# METHOD FOR REDUCING OF NOISE BY IMPROVING SIGNAL-TO-NOISE-RATIO IN WIRELESS LAN

Dr.R.Seshadri [1] and Prof.N..Penchalaiah [2]

[1]Prof & Director of university computer center, S.V.University, Tirupati, India
ravalaseshadri@gmail.com
[2]Department of Computer Science Engineering, ASCET, Gudur, India
pench_n@yahoo.com

## ABSTRACT

The signal to noise ratio (SNR) is one of the important measures for reducing the noise.A technique that uses a linear prediction error filter (LPEF) and an adaptive digital filter (ADF) to achieve noise reduction in a speech and image degraded by additive background noise is proposed. Since a speech signal can be represented as the stationary signal over a short interval of time, most of speech signal can be predicted by the LPEF. This estimation is performed by the ADF which is used as system identification. Noise reduction is achieved by subtracting the reconstructed noise from the speech degraded by additive background noise. Most of the MR image accelerating methods suffers from degradation of acquired images, which is often correlated with the degree of acceleration. However, Wideband MRI is a novel technique that transcends such flaws.In this paper we proposed LPEF and ADF for reducing the noise in speech and also we demonstrate that Wideband MRI is capable of obtaining images with identical quality as conventional MR images in terms of SNR in wireless LAN.

## 1. INTRODUCTION

In recent years, research on methods of noise reduction in a speech degraded by additive background noise is actively being done by the use of microphone array [1], spectrum subtraction (SS) [2], etc. Imperfection can be seen in the method of the noise reduction using two microphones which can be considered as a directional microphone with a blind spot in the arrival bearing of the noise. When many noise sources exist, an increase in number of microphones cannot be avoided. It is therefore important to develop a noise reduction method which uses a single microphone, and which can cancel multiple noise sources. In the systems with only one microphone, extracting a speech &om a speech degraded by additive background noise requires the use of SS method. One of the SS methods [2] improves the signal to noise ratio(SNR) at the expense of processing delay, signal distortion and musical tones that arise due to the residual noise. Moreover, SS method needs an advance estimation of noise spectrum. It means that the SS method requires voicehoiceless section detector under the practical environment. In order to improve on these negative effects, we have investigated a noise redudon method based on linear prediction [3]. In this method, a noise reduction is efficiently performed for additive white noise, because the coefficients of the linear predictor converge such that the prediction error signal becomes white [4]. However, when a background noise is colored, effectiveness of the noise reduction decreases since the linear predictor estimates both a speech and a colored noise spectrum.

It is known that the coefficients of the LPEF converge such that the prediction error signal becomes white [4]. Since a speech signal can be represented as the stationary signal over a short interval of time, most of speech signal is predicted by the LPEF. On the other hand, when the input signal of the LPEF is a background noise, the prediction error signal becomes white. Assuming that the background noise is generated by exciting a linear system with a white noise, then we can reconstruct the background noise from the prediction error signal by estimating the transfer function of noise generation system. This estimation is performed by the ADF which is used for system identification. Noise reduction is achieved by subtracting the reconstructed noise from the speech degraded by additive background noise.

In this paper, we propose a technique that uses a linear prediction error filter (LPEF) and an adaptive digital filter (ADF) to achieve noise reduction in a speech in wireless LAN.

The concept of Wideband MRI is based on the multi-carrier modulation technique in wireless communications and it increases the bandwidth in MRI. Wideband MRI simultaneously excites and acquires multiple slices using signals with multiple frequencies. RF pulses in Wideband MRI contains several bands, we define the number of bands as "Wideband multi-slice/slab factor W". It is the excitation/acquisition of this wideband signal that provides additional information necessary for acceleration. We have reported Wideband MRI acceleration technique and its potential previously [1]. Several promising applications can be realized with the help of this technique and we have demonstrated them successfully [2,3,4]. Nevertheless, image properties of the Wideband MRI technique have to be thoroughly examined in order to ensure its feasibility and validity. In recent years, several proposals of using wireless transmission for MRI have been reported to avoid the interference between array channels [1-3]. Amplitude modulation (AM) and single sideband (SSB) analog wireless techniques have been applied to design transponders for RF coils [4].Compared to this analog transmission technique, digital transmission has advantages of better noise immunity, more stability and flexibility, and is code error free. In this work, we have designed and implemented a digital transmission system based on WLAN 802.11b standard, which can reach the speed of 11Mbps with 2.4G band.

## 2. THE PROPOSED NOISE REDUCTION METHOD

In this section, we describe the principle of the proposed noise reduction method using the LPEF and the ADF. Below equation shows a transversal type LPEF, where x(n) is a speech v(n) degraded by background noise e(n)= x(n)-y(n) is the prediction error signal, and An) is the predicting signal, y(n) is given by

$$y(n) = \sum_{k=1}^{M} h_k(n) x(n-k) \qquad (1)$$

where hk(n) (k1,2, ..., M) are the tap coefficients [4]. The coefficients of the LPEF will converge such that the prediction error signal e(n) becomes white. When small step size is used in the algorithm for updating the coefficients of the LPEF, the LPEF estimates the input signal with high fidelity at the expense of poor tracking ability. On the other hand, when the large step size is used in the algorithm, the LPEF estimates the input signal with high tracking ability at the expense of roughly estimating. Therefore, it is required to whiten the input signal x(n) that the large step size is used in the algorithm for whitening the non-stationary speech signal, and the small step size is used in the algorithm for whitening the noise signal. Specifically, the algorithm with large step size is used for updating the LPEF's coefficients corresponding to the samples around the sample delayed by a pitch period. The other coefficients of the LPEF are updated by using the algorithm with small step size. Then the LPEF estimates the voiced speech with high tracking ability, and estimates the other signal with high fidelity. Thereby, the signal whitening by the LPEF is efficiently performed. Next, we consider the reconstruction of the background noise from the prediction error signal.

Assuming that the background noise is generated by exciting a linear system with a white noise as shown in Fig. 2, then we can consider the system identification model as shown in Fig. 3(a) and (b), where H,(z) represents the transfer function of noise generation linear system, HADF(z) is the transfer function of the ADF, hi(n),h,'(n), ..., hi(.) are the tap coefficients of the ADF,&n is the reconstructed noise, and $e_{ADF}(n) = \hat{v}(n)$;(n) is the extracted speech. The ADF cannot estimate the v(n) due to the e(n) which does not correlated with the voiced speech v(n). Then the output of the ADF represents only the background noise. However, when the unvoiced speech whose source is represented as random noise is observed in the v(n), a probability of reconstructing the unvoiced speech by the ADF arises. In order to improve this problem, we use the long time average of the error signal e,"() for updating the coefficients of the ADF since that the unvoiced speech vanishes in a short time in comparison to the noise. Thereby, the probability of

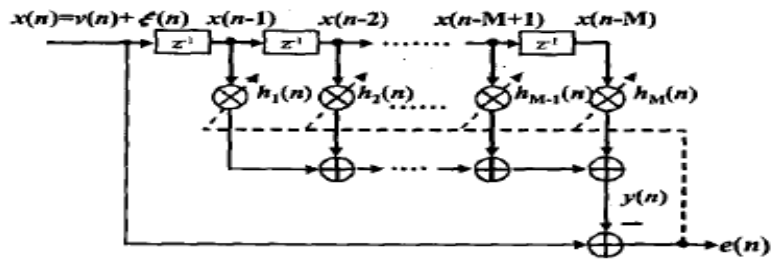

Figure 1.    LPEF (Linear Prediction Error Fileter)

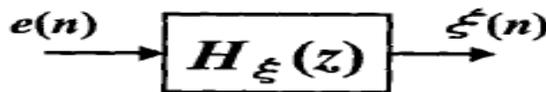

Figure 2.    Noise generation system

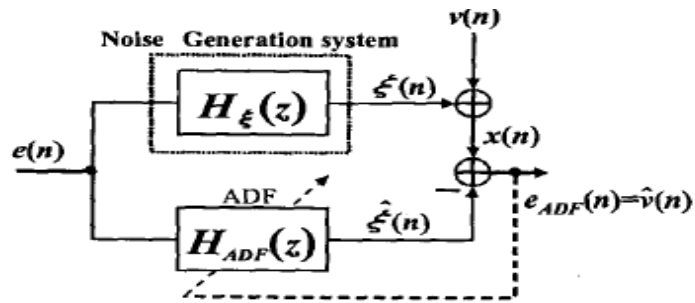

(a) Block diagram of System identification model

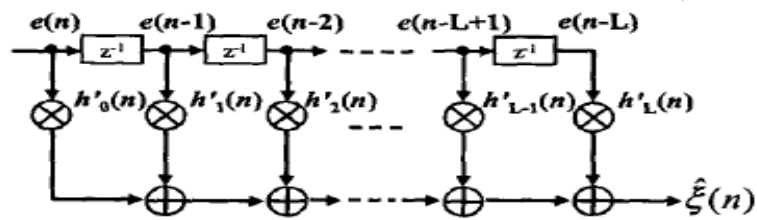

(b) Structure of ADF (Adaptive Digital Filter)

Figure 3.    System Identification Model

Reconstructing the unvoiced speech will be decreased. That is, we apply the small step size to the adaptive algorithm for updating the coefficients of the ADF. We shall now incorporate the signal whitening Process with the noise reconstruction process for noise reduction. The proposed noise reduction system is shown in Fig. 4, where the &.EF(z) represents the transfer function of the LPEF. The noise reduction is achieved by subtracting the reconstructed noise ( n ) from the input signal x(n).

# 3. MATERIALS AND METHODS

To compare the image quality with and without wideband MRI, two identical standard Broker head phantoms were used. Each phantom has several structures for image quality analysis. There are 1.4mm and 1mm wide horizontal and vertical stripes specifically tailored for image resolution assessment while some cylindrical structures containing oil and porous materials can be used for contrast evaluation. All images were acquired on a 3T Brokers Biopsies MR imaging system, with the use of 8826 head coil. The two phantoms were placed 14cm apart along the axial direction in the RF coil. Axial images of the two phantoms were first taken separately without using the Wideband MRI technique. Then a Wideband factor of W=2 was applied by a modified Sinc RF pulse to obtain simultaneously the images of both phantoms using half the total scan time. Imaging parameters are listed as below: FOV=19cmx19cm, total Matrix size=256x256, gradient echo sequence with TR/TE=30/6.3ms. To analyze the image, SNR signal was sampled from 9 uniform areas throughout the image and noise was calculated as the standard deviation at four corners of the image.

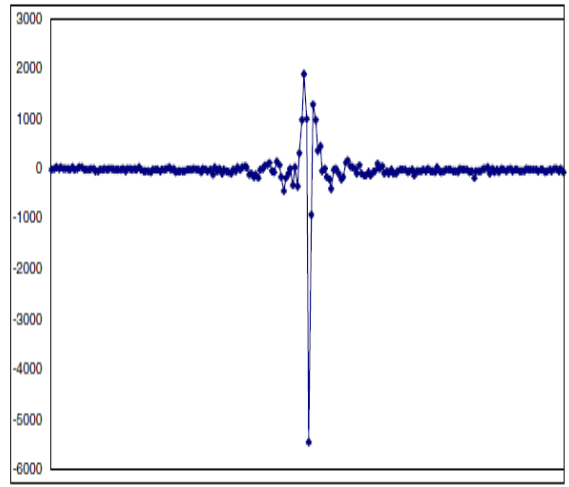

The complete images need both real and imagery parts. However, imaginary parts of imaging signals can not be generated just through some simple algorithms after AD conversion. Because it is difficult to decide the sampling rate of sine wave and cosine wave in one cycle to make real and imaginary signals synchronous. The synchronous imaginary part of MR imaging signals could be completed in hardware for the real part circuit simultaneously.

## 4. CONCLUSION

We have proposed a new noise reduction method using linear prediction error filter and adaptive digital filter. From the experimental results, it was observed that there was improvement of SNR in the extracted voice signal, and this proposed noise reduction method is also available under the practical environment. Further researches involve an improvement of the tracking ability to the non-stationary noise like a tunnel noise, a reduction of the residual noise and a performance evaluation in a test product. We believe it is self-evident that Wideband MRI is a powerful MRI accelerating technique that can maintain the same image quality of each accelerated image while other acceleration methods suffer from degradation such as SNR loss or artifacts. Since Wideband MRI accelerates by the increase in bandwidth instead of k-space or image space alteration, properties stated in this study can be further extended into higher Wideband accelerated images.

**Authors:**

**Dr.R.Seshadri** Working as Professor & Director, University Computer Centre, Sri Venkateswara University, Tirupati. He was completed his PhD in S.V.University in 1998 in the field of " Simulation Modeling & Compression of E.C.G. Data Signals (Data compression Techniques) Electronics & Communication Engg.". He has richest of knowledge in Research field, he is guiding 10 Ph.D in Fulltime as well as Part time. He has vast experience in teaching of 26 years. He published 10 national and international conferences and 15 papers published different Journals. 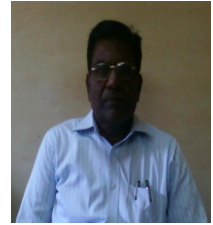

**Prof.N.Penchalaiah** Research Scholar in SV University, Tirupati and Working as Professor in CSE Dept,ASCET,Gudur. He was completed his M.Tech in Sathyabama University in 2006. He has 11 years of teaching experience. He guided PG & UG Projects. He published **2** National Conferences and **6** Inter National Journals. 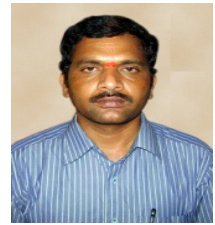